\documentclass[usenatbib]{mnras}
\usepackage[utf8]{inputenc}
\usepackage{amsmath}
\usepackage{graphicx}
\usepackage[T1]{fontenc}
\usepackage{ae,aecompl}

\title[Effect of angular momentum on density profiles]{Forming disc galaxies in major mergers: III. The effect of angular
  momentum on the radial density profiles of disc galaxies}

\author[N. Peschken, E. Athanassoula, S. A. Rodionov]{N. Peschken$^{1,2}$\thanks{Contact e-mail: \href{mailto:npeschken@camk.edu.pl}{npeschken@camk.edu.pl}}, E. Athanassoula$^{1}$, S. A. Rodionov$^{1}$
\\
$^{1}$Laboratoire d'Astrophysique de Marseille, 38, rue Frédéric
Joliot-Curie 13388 Marseille cedex 13 FRANCE \\
$^{2}$Nicolaus Copernicus Astronomical Center, Polish Academy of Sciences, ul. Bartycka 18, 00-716 Warsaw, Poland}

\pubyear{2017}

\begin{document}
\label{firstpage}
\pagerange{\pageref{firstpage}--\pageref{lastpage}}
\maketitle

\begin{abstract}
We study the effect of angular momentum on the surface density 
profiles of disc galaxies, using high resolution simulations of major mergers whose remnants have downbending radial density profiles (type II). As described in the previous papers of this series, in this scenario, most of the disc mass is acquired after the collision via
accretion from a hot gaseous halo. We find that the inner and outer
  disc scalelengths, as well as the break radius, correlate with the total  
angular momentum of the initial merging system, and are larger for high angular momentum systems. We follow the angular momentum 
redistribution in our simulated galaxies, and find that, like the mass, the disc angular momentum is acquired via accretion, i.e. to the detriment of the gaseous halo. Furthermore, high angular momentum systems give more angular momentum to their discs, which affects directly their radial density profile.
Adding simulations of isolated galaxies to our sample, we find that the correlations are valid also for disc galaxies evolved in isolation. 
We show that the outer part of the disc at the end of the simulation is populated mainly by 
inside-out stellar migration, and that in galaxies with higher angular momentum, stars travel 
radially further out. This, however, does not mean that outer disc stars (in type II discs) were 
mostly born in the inner disc. Indeed, generally the break radius increases over time, and not taking this into account leads to overestimating the number of stars born in the inner disc.
\end{abstract}


\section{Introduction}

The early pioneering work of \cite{Freeman.70} showed clearly that the radial surface density profile of disc galaxies is well fitted by an exponential. Later work (\citealt{Kruit.79}, and later e.g. 
\citealt{Pohlen.DLA.02}; \citealt{Erwin.BP.05}; \citealt{Gutierrez.EAB.11}) 
revealed the presence of a break in 
the profile of most galactic discs, as well as the fact that both the 
inner and the outer parts are well described by exponentials.

The break can be of two kinds, depending on the scalelengths of 
the two exponentials. If the slope of the outer part of the disc, hereafter called 
outer disc, is steeper than the slope of the inner part (inner disc), i.e. if the inner disc scalelength is greater than the outer disc one, the profile 
is called downbending, or type II (\citealt{Pohlen.Trujillo.06}). This
is the most common profile for disc galaxies (e.g. \citealt{Pohlen.Trujillo.06}; \citealt{Azzollini.TB.08}; \citealt{Laine.LS.14}). In the opposite case, the outer disc
is shallower than the inner disc, and the profile is called upbending, 
or type III. The single exponential case, where no break is observed, is called type I.

\noindent Several mechanisms producing those different types of discs have been proposed so far. While it has been argued that a single exponential is the canonical profile for a 
disc (\citealt{Gunn.1982}; \citealt{Lin.Pringle.1987}; \citealt{Yoshii.Sommer-Larsen.89}; 
\citealt{Ferguson.Clarke.01}; \citealt{Elmegreen.Struck.13, Elmegreen.Struck.16}; \citealt{Herpich.TR.16}; \citealt{Struck.Elmegreen.17}), many formation scenarii have been proposed for the downbending discs, sometimes related to each other.
They are thought to be created by bars via the Outer Lindblad Resonance (\citealt{Debattista.MCM.06}; \citealt{Pohlen.Trujillo.06}; 
\citealt{Munoz-Mateos.SG.13}; \citealt{Kim.GSA.14}), by star 
formation thresholds (\citealt{Schaye.04}; 
\citealt{Pohlen.Trujillo.06}; \citealt{Elmegreen.Hunter.06}), or 
connected with morphological components such as rings or spirals (\citealt{Laine.LS.14}).
Stellar migration is also found to be important in the creation of type II profiles, being 
able to redistribute stars from the inner disc to the outer disc,
and can be coupled with a star formation threshold 
(\citealt{Roskar.DSQ.08}). On the other hand, type III profiles 
(upbending discs) remain poorly understood. They are sometimes associated 
with a spheroidal component such as a stellar halo, or with the superposition of a thin and a 
thick disc, and could be the result of minor 
mergers, or be linked to a strong bar (Erwin et al. 2005; \citealt{Younger.CSH.07};
\citealt{Bakos.Trujillo.12}; \citealt{Comeron.ES.12}; \citealt{Herpich.SRM.15}).\\

\noindent Although the majority of the present-day spirals is thought to have 
experienced at least one major merger in their history 
(e.g. \citealt{Hammer.FPY.09}), so far most of the simulated galaxies used to  
investigate the formation of these different disc types 
 have been evolved in isolation. Recently, however,
 we presented in \citet[hereafter A16]{Athanassoula.RPL.16} three fiducial  
examples from a 
large sample of high resolution simulations of a major merger between two disc galaxies with a hot 
gaseous halo each, and showed that the remnants are good models of 
spirals. In this paper, we aim to study the role of angular momentum in shaping type II profiles, obtained from our large sample of major merger 
simulations.

The outline is as follows. In section 2, we briefly summarise the 
necessary parts of A16, and describe the fitting procedure for the 
radial density profiles. We compute the initial angular
  momentum and link it to the scalelengths in section 3.  
In section 4, we follow the angular momentum redistribution to explain how the initial angular momentum can affect the final  
disc properties. We discuss our results in section 5, and 
conclude in section 6.

\section{Technical aspects}
\label{sec:tech}

\subsection{Description of the simulations}
\label{sec:desc_simus}

In this section, we will briefly summarise the simulations
characteristics, which have been obtained and described in some detail in A16. 
Our simulations start from two spherical protogalaxies consisting only of dark
matter (DM) and hot gas, which we set on a given orbit. We use a total of 5.5 million particles (2 million for the baryons, 3.5 million for the DM), with
softenings of 25 pc for the gas and 50 pc for the DM. Each particle's
mass is 5$\times$10$^4$ M$_{\odot}$ for the gas and stars, and 2$\times$10$^5$
M$_{\odot}$ for the dark matter. By the time of the merging, a disc
has formed in each of the progenitors, which is destroyed
by the merging, its stars ending up mainly in a classical bulge. Gas
continues to fall from the halo, and new discs, both thin and thick, are gradually formed in
the remnant. Well-defined spiral arms soon develop in the thin disc, as
well as a bar and a boxy/peanut bulge. Each simulation ends after 10
Gyr evolution, showing a remnant with a classical bulge-to-total ratio
which is consistent with that of real spiral galaxies.

Our simulations are made using the N-body/SPH code GADGET3, including gas
and its physics. The description of this code can be found in
\cite{Springel.Hernquist.02} and \cite{Springel.05}. Stars and dark matter are modeled 
  by N-body particles, and gas by SPH particles, with fully adaptive 
  smoothing lengths. 
  Gravity is computed with a 
  hierarchy tree algorithm, and the code uses subgrid physics for the 
  feedback, star formation and cooling, described in 
  \cite{Springel.Hernquist.03}.
  
For a description of the technical aspects of the 
simulations we refer the reader to \citet[hereafter R17]{Rodionov.AP.17}. To avoid an excessive central concentration in our
  simulated galaxies, which would lead to unrealistic circular velocity
  curves and would delay the formation of the bar, we added AGN feedback (R17). This is based on a density threshold $\rho_{AGN}$ and
  a temperature $T_{AGN}$, while the physics underlying it is described in
  detail in A16.
  
The simulations have a variety of different orbits for the two
merging protogalaxies. These orbits are characterised by their
ellipticity, and the initial distance between the two
progenitors. Each orbit leads to a different merging time, which is
difficult to define precisely, but can
be approximated by deriving the time beyond which the distance between the two centers of
density stays below 1 kpc (see A16).

The two protogalactic haloes start with an initial spin, characterised by a
spin value $f$ representing the fraction of particles rotating with
a positive sense of rotation. Thus for $f=1$, all the
particles rotate in the direct sense, while for $f=0.5$ there is no net rotation. \\

\noindent The simulations used in this paper have rather early merging times, i.e. between 1.2 and 2.2 Gyr after the start of the simulation. About two thirds of our simulations have spin axes
perpendicular to the orbital plane, but for the other simulations we tilt one or both
protogalaxies by a chosen angle, to see the effect of the spin axis
orientation. In most simulations the spin value $f$ is the same for both protogalaxies, but in some simulations we introduced a different spin in each protogalaxy. We ran simulations of mergers with mass ratios between the two protogalaxies of 1, 1/2, 1/3, 1/4 and 1/8.

A central AGN is present in the remnant galaxy of the majority of our simulations,
as described in A16 and R17, but we include 21 simulations without AGN to cover a larger part of the available parameter space. Nevertheless, the presence
of our AGN affects mainly the central part of the disc (R17), and thus should not have an impact on
the analysis presented in this paper.

We will first consider a subsample of 132 simulations which all have the same total mass, i.e. the same number of particles (5.5 million), and call it sample A.
We also ran 67 simulations with various masses, each different from the mass of the simulations in sample A. This is done by keeping the same mass for single particles, but changing the number of particles. We add this group of 67 simulations (called sample B) to sample A to obtain a new sample of 199 simulations, sample A+B, which we will use in the discussion (section \ref{sec:diffmtot}).

\begin{figure}
\centering
\includegraphics[scale=0.3]{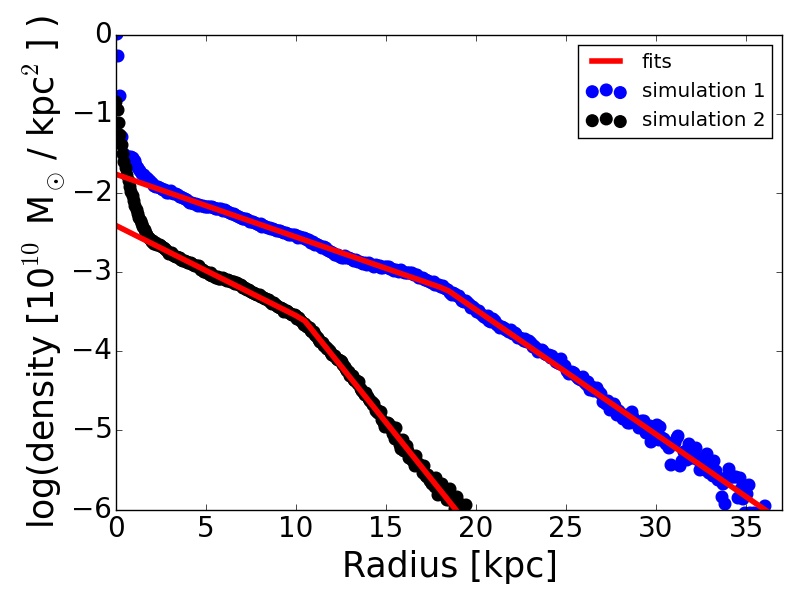}
\caption{Projected surface stellar density radial profiles for two
  simulations, together with the corresponding fits for the disc
  part. For clarity, the second simulation (in black) has been shifted
  down by 1 logarithmic unit. Simulation 1 has a higher initial spin
  parameter $\lambda$ than simulation 2.}
\label{fit}
\end{figure}

\subsection{Fitting the radial density profiles}
\label{sec:fit}

To derive the stellar radial density profile, we use axisymmetric concentric cylindrical
annuli, and choose z$_{lim}$=1 kpc as maximum height to keep only the thin disc profile. 
As shown in A16, the disc at the end of our three fiducial simulations is composed of an inner and  
  a downbending outer disc (type II), separated by a break. This is
  also the case for the 199 simulations of the sample used in this
  paper (sample A+B), therefore we fit the disc part of our  profiles with two
exponential functions. We use a
``piecewise'' fit, which means that we fit the inner and the outer
disc separately with an exponential:

\begin{gather}
\Sigma_{in}(R)= \Sigma_{i} \ \exp(\frac{-R}{h_i}), \hspace{1cm}
R<R_{break} 
\\
\Sigma_{out}(R)= \Sigma_{o} \ \exp(\frac{-R}{h_o}), \hspace{1cm} R>R_{break}
\end{gather}

\noindent where  $\Sigma_{i}$, $\Sigma_{o}$ are normalization
factors, $h_i$ and  $h_o$ are the inner and outer disc scalelengths, and
$R_{break}$ the break radius. The latter is derived by taking the intersection between the inner and
the outer disc fits. The interval to fit for each part of the disc is
selected manually, and we exclude, after visual inspection, 
every simulation for which the fit is not reliable. Two examples of fits are shown in Fig.
\ref{fit}. Another way to make this fit would be to use a double
exponential function (or ``broken-exponential'') for the whole disc, which has been shown to give
very similar results (differences in the scalelengths $<$ 5 per cent,
Erwin et al. 2008). We chose the present approach because of its simplicity.

The scalelengths and the break radius change with time, as shown both by observations (\citealt{Perez.04}; Azzolini et al. 2008), and simulations (\citealt{Roskar.DSQ.08}; Athanassoula, Peschken \& Rodionov, in prep, hereafter Paper IV), and to be able to compare
consistently the scalelengths derived for the different simulations,
we need to compare all galaxies at the same evolutionary time, i.e. at the same time after the merging. We thus take a fixed evolution time of 7.8 Gyr after the
merging for every simulation. Given that the merging times have
values between 1.2 and 2.2 Gyr, the times at which we now study the
disc are therefore between 9 and 10 Gyr, depending on the
simulation. We will hereafter refer to this time as the final
state. \\

\noindent We here aim to investigate the role of angular momentum in the formation of the new disc formed after the merging, in the secular evolution period of the galaxy. 
Nevertheless, some of the stars 
formed before or during the merging were spread all over the
 galaxy by the merging, and can pollute our density  
profiles. These stars 
represent between 10 and 30 per cent of the total stellar mass in the thin
 disc. Most of them are located in the bulge  
region (about 65 per cent), but they can also 
form a stellar halo or a thick disc, which will add particles 
to the thin disc, changing the  
value of the scalelengths by up to
 20 per cent. These stars are not related to the disc formation and its dependance on the angular momentum, and we thus chose  
to remove them, selecting only the stars formed
 after the merging
 to derive our density profiles and the corresponding scalelengths.

\section{The role of angular momentum}
\label{sec:main}
\subsection{Definition of the global, dimensionless spin parameter}
\label{sec:lambdadef}

Here, we use the dimensionless spin parameter $\lambda$ (\citealt{Peebles.69}):

\begin{equation}
\centering
\lambda=\frac{L|E|^{1/2}}{GM^{5/2}},
\end{equation}

\noindent where $L$, $E$ and $M$ are respectively the total angular momentum, energy and mass of
the system including its dark matter, gas and stars, computed with respect
to the center of mass of the system, and $G$ is the
gravitational constant. The total energy
is computed adding the kinetic energy, the gravitational potential
energy and the gas internal energy.
The total angular momentum $L$ is a conserved quantity, but the total energy varies with
time because of processes such as the cooling of the gas and the
stellar feedback. We therefore have to specify the time at which we will calculate $\lambda$, and make sure it is calculated consistently for all simulations. To avoid the technical difficulties inherent in using the merging time, we use t=1 Gyr before
the merging, which we call the initial state.

\subsection{Angular momentum and scalelengths}
\label{sec:corr}

\begin{figure}
\centering
\includegraphics[scale=0.27]{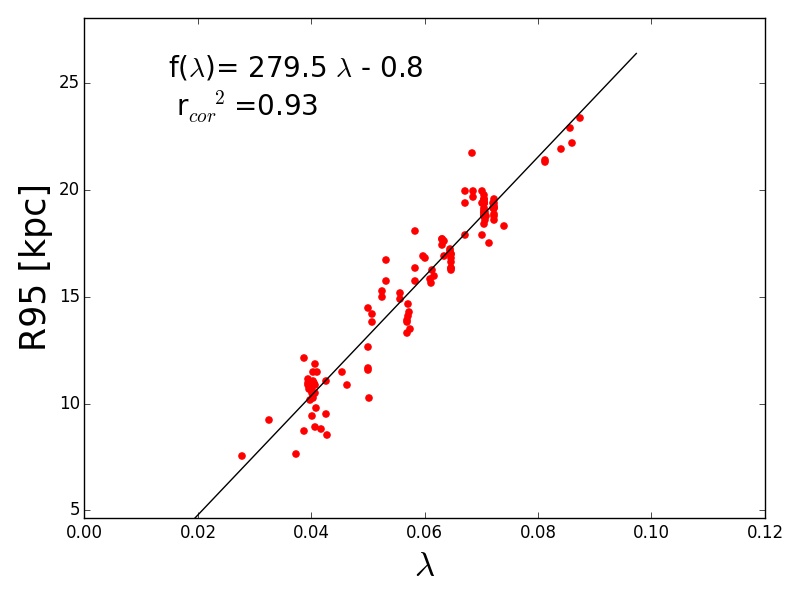}
\caption{Cylindrical radius containing 95 per cent of the stellar mass at the final state, as a function of the spin parameter $\lambda$ computed at the initial state, for the simulations of sample A. The corresponding linear fit is plotted in black, with the equation and correlation coefficient given in the top left corner.}
\label{r95}
\end{figure}

How does the disc structure relate to the global spin parameter
  $\lambda$? From the definition of angular momentum, we
  expect that for two galaxies of the same total mass,
  the galaxy with the highest angular momentum will be more
  extended, as it has been shown in previous studies (e.g. \citealt{Dalcanton.SS.97} and \citealt{Kim.Lee.13}). We checked this result in our simulations by plotting the size of our galactic discs at the final state as a function of $\lambda$, for the simulations of sample A. As an estimate of the disc size, we take $R95$, the cylindrical radius containing 95 per cent of the total stellar mass.  We can see on Fig. \ref{r95} that the size of the final galaxy increases linearly with $\lambda$, with a high correlation coefficient.

\begin{figure}
\centering
\includegraphics[scale=0.3]{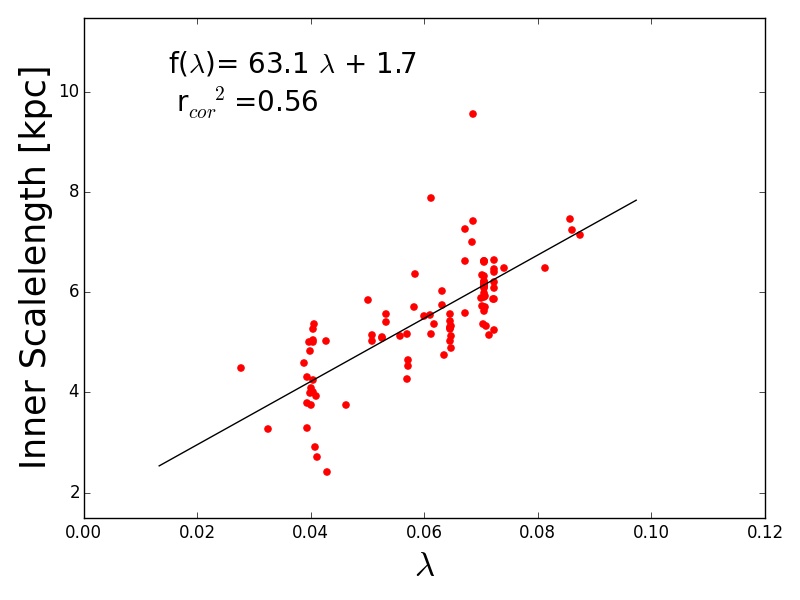}
\includegraphics[scale=0.3]{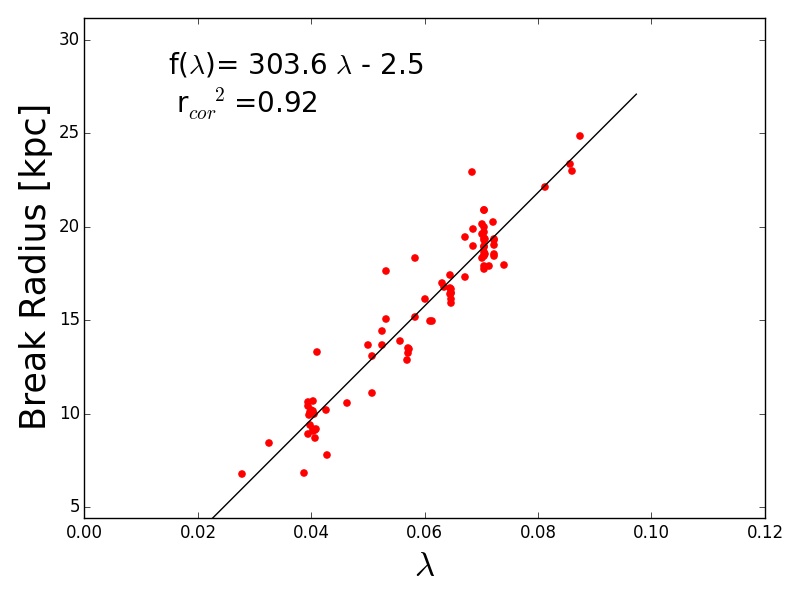}
\includegraphics[scale=0.3]{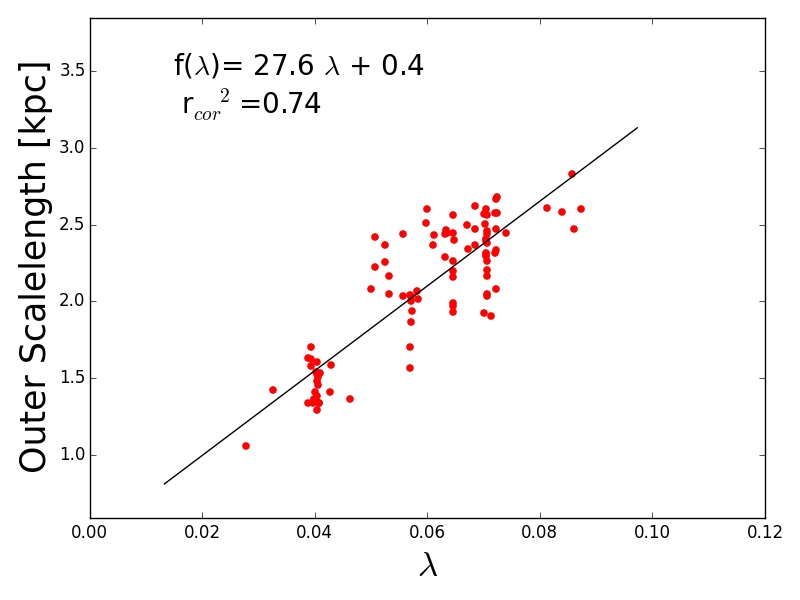}
\caption{From top to bottom we plot respectively the inner
  scalelengths, the break radii and the outer scalelengths derived from
the fits at the final state, each as a function of the spin
parameter $\lambda$  taken at the initial state, for the simulations in sample A. The corresponding linear fits are plotted in black, with the equation and correlation coefficient given in the top left corner of each plot.}
\label{res}
\end{figure}

\begin{figure*}
\centering
\includegraphics[scale=0.2]{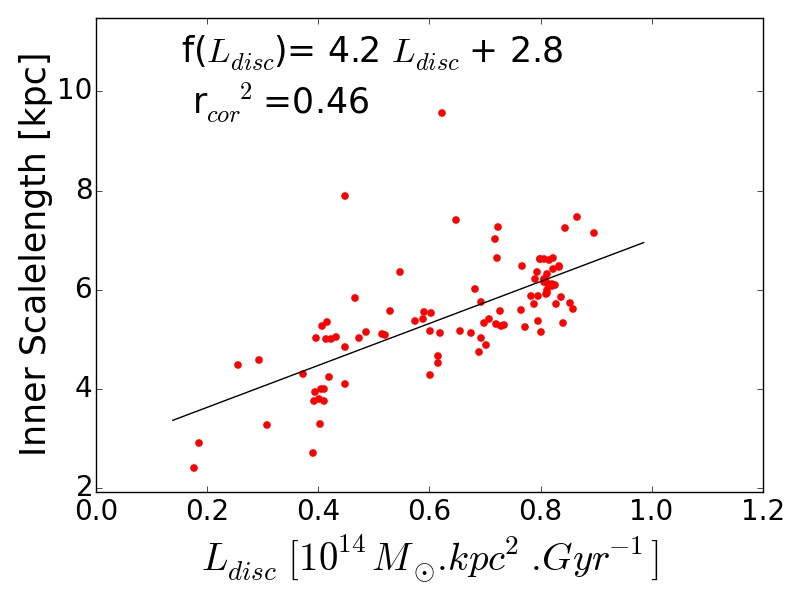}
\includegraphics[scale=0.2]{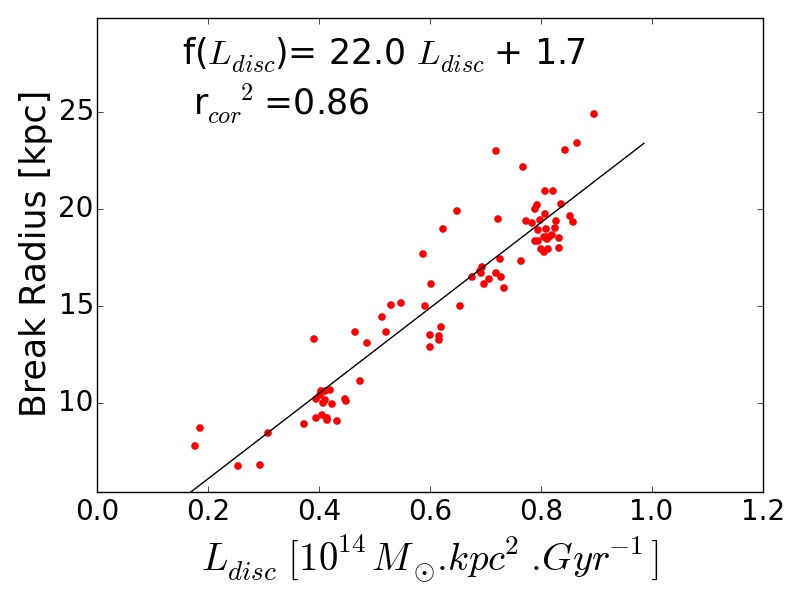}
\includegraphics[scale=0.2]{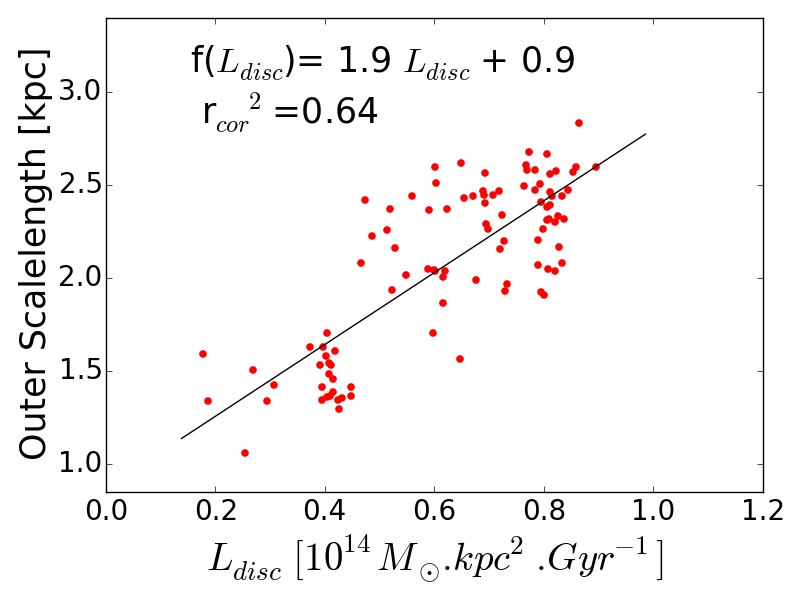}
\caption{Correlations of the inner scalelength, the break radius and the outer scalelength, with the total baryonic angular momentum of the final disc, for sample A.}
\label{Jcorr}
\end{figure*}

  We thus find the disc to be globally larger for high $\lambda$
  galaxies, as expected. However, since the disc is constituted of an inner and an outer
  part, which part of the disc is affected most? We expect
  at least one of the two discs (inner 
  or outer) to be larger at higher $\lambda$.
To have a first insight of the effect of the spin parameter $\lambda$ on the parameters of the radial density
profiles, we plot in Fig. \ref{fit} the profiles at the final state (as defined in section \ref{sec:fit}) for two simulations of sample A with different $\lambda$ values. We can
see that the scalelengths of the simulation with a higher $\lambda$ (in blue) are larger, and its break is located further out. The next step is to see whether this preliminary result is valid for all the simulations in sample A, as described below.

We examine the discs of the remnants at the final state, and plot the three parameters derived from the fits of the
radial density profiles (inner, outer disc
scalelength and break radius) as a function of the initial
$\lambda$ for each simulation in sample A (Fig. \ref{res}). As described in
section \ref{sec:fit}, we excluded all the simulations for which there was some uncertainty in
the fit of the corresponding parameter, which explains why the number
of points is different for each plot. We thus have a total of
97 values for the inner disc scalelength (hereafter inner scalelength), 99 for the outer
disc scalelength (outer scalelength) and 84 for the break radius (for which the fits of both the inner
and the outer disc have to be reliable).

We see that the inner scalelength, the outer scalelength and the break
radius \textit{all}
increase linearly with $\lambda$, so that larger angular momentum
systems produce discs with larger inner, outer scalelengths and break
radii. This confirms the results of \cite{Herpich.SD.15} for the
inner scalelength and the break radius for cases with isolated galaxies with no mergers (see discussion in section
\ref{sec:discu_gen}), and adds a further argument showing that the remnant of a major merger possesses all
the characteristics of a disc galaxy.

The correlations of the fits in Fig. \ref{res} are tight, especially for the break
radius. The scalelengths show a larger spread, presumably because their values
can be more strongly influenced by the presence of morphological components such as
bars, spirals or rings than the break radius. \\

\noindent It is important to note that all the simulations we have so far considered (sample A) have identical total masses (gas + stars + DM) and similar
total stellar masses at their final state, so that the spread of
scalelengths cannot be simply due to a mass dependence. Thus the spread should be due to the different parameters
  of the initial conditions in the
various simulations, such as the halo spin value, and, in particular, to the different orbits of the two progenitors.

\section{Understanding our correlations}
\label{sec:understand}

The dimensionless spin parameter $\lambda$ is computed 1 Gyr before the merging, and is dominated by the halo (hot gas and DM), because the discs of  
the protogalaxies are of short extent and of low mass, as the merging occurs early in our sample of simulations. Therefore, $\lambda$ does not seem to be directly related to the final disc of the remnant and to its properties such as the scalelengths. To understand how it can yet influence the final disc, we will first look at the angular momentum in the final disc. 

\subsection{Angular momentum in the baryonic disc}
\label{sec:bary_disc}

We define the baryonic disc as the 2 kpc thick pill-box ($|z|<1$ kpc, consistently with section \ref{sec:fit}) of maximum radius $R_{max}=1.5 \ R_{break}$, and exclude the inner 3 kpc to remove the bulge part. We then compute the total angular momentum of the baryonic matter (gaseous + stellar particles) in this volume at the final state,  
and plot it against the scalelengths and the break radius for all the simulations of sample A in Fig. \ref{Jcorr}. We find clear correlations, showing that the properties of the disc are directly linked to its angular momentum.

We now need to relate this disc angular momentum to the initial halo-dominated spin parameter $\lambda$.
We showed in A16 that the gas from the gaseous halo accretes onto the
disc and thus rebuilds it after the merger, by fuelling 
star formation in the disc throughout the whole simulation. We suggest here that gas accretion from the halo is the main  
mechanism gradually transferring angular momentum from the gaseous halo to the disc, and investigate this in the following subsections.

\subsection{Angular momentum redistribution}
\label{sec:redis}

To be able to study the exchanges of angular momentum between the disc
and the gaseous halo, we first made sure that the
angular momentum exchanges between the baryonic and the dark matter
haloes are relatively small and thus can be neglected in the context of
our very simple qualitative explanation. Indeed, we found such
exchanges to be of the order of few per cent.\\

\noindent To illustrate the angular momentum exchanges in our simulations, we will focus 
on two fiducial simulations of sample A, $mdf732$ and $mdf780$, presented in A16,
which have quite different $\lambda$ values, $mdf732$ having a
considerably higher $\lambda$ than $mdf780$. Note, however, that we
made the same analysis with other simulations as well, and  
found similar results.

\noindent We define the disc as in section \ref{sec:bary_disc}, and the halo as everything else. The baryonic halo is mainly constituted
of gas, but a few stars are expected to be found as
well. Nevertheless, we find the total angular momentum $L$ of these
stars to be a thousand times lower than the halo gas, and we thus neglect them.

\begin{figure}
\centering
\includegraphics[scale=0.3]{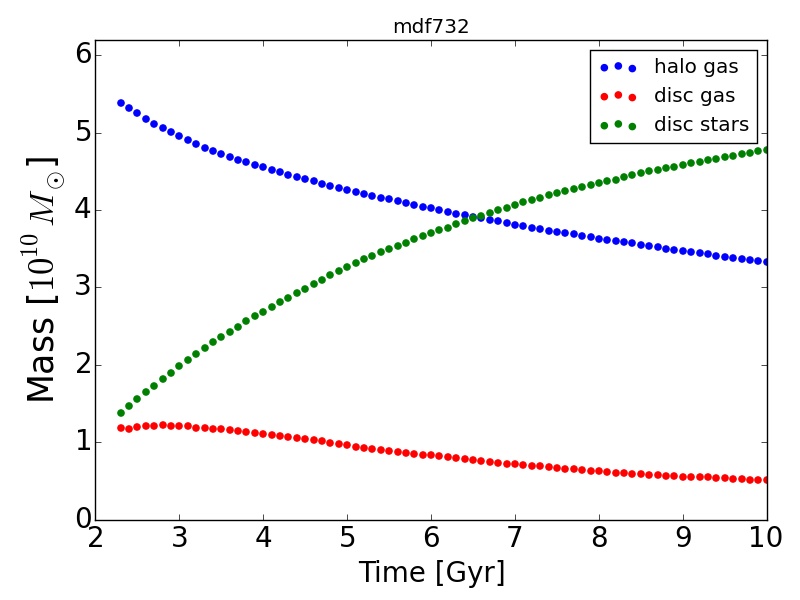}
\includegraphics[scale=0.3]{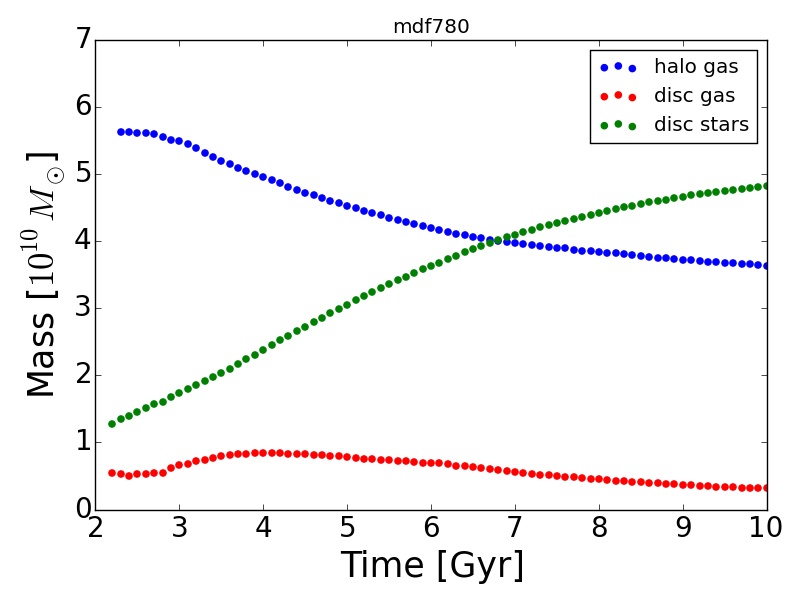}
\caption{Total mass of the gaseous halo, the gaseous disc and the stellar disc as a function of time for two simulations, mdf732 (high $\lambda$, upper pannel) and mdf780 (low $\lambda$, lower pannel).}
\label{M_all}
\end{figure}

\begin{figure}
\centering
\includegraphics[scale=0.3]{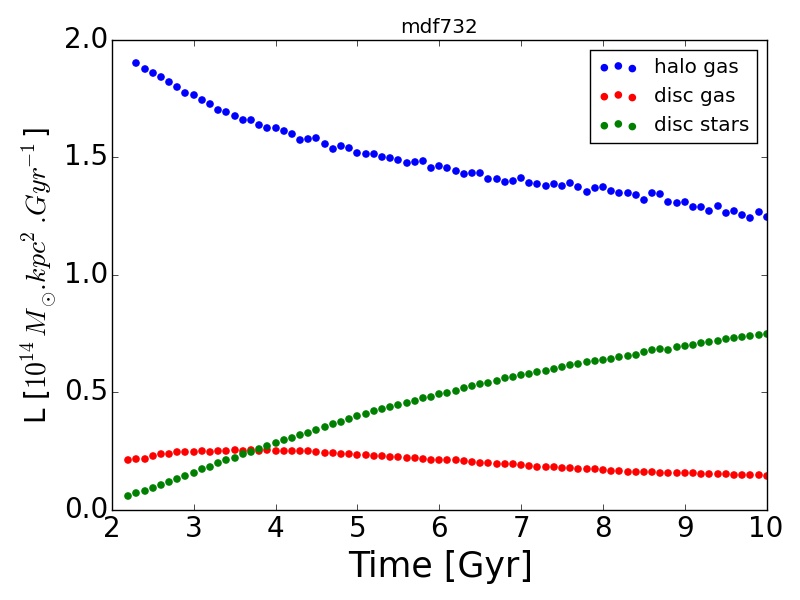}
\includegraphics[scale=0.3]{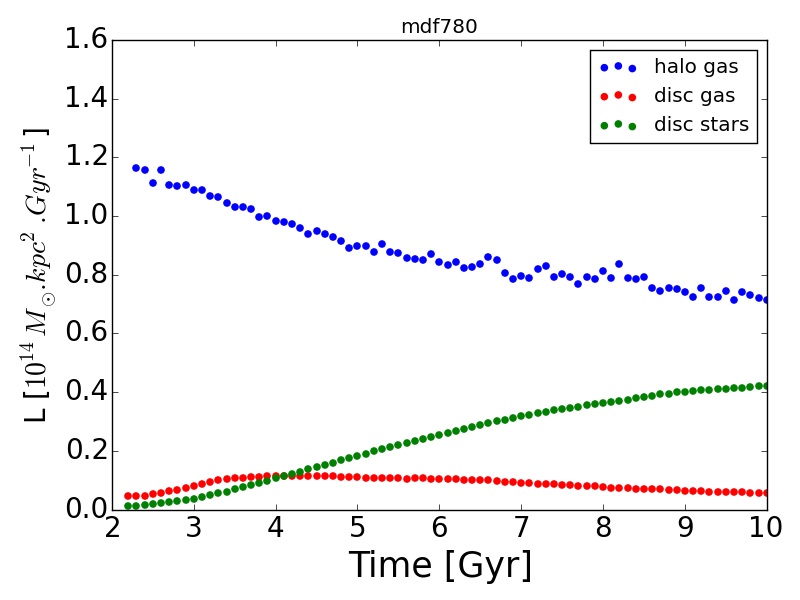}
\caption{Total angular momentum of the gaseous halo, the gaseous disc and the stellar disc as a function of time for two simulations, mdf732 (high $\lambda$, upper pannel) and mdf780 (low $\lambda$, low pannel).}
\label{J_all}
\end{figure}

We first plot in Fig. \ref{M_all} the evolution over time of the total mass of the halo gas, the disc gas and the stars in the disc, separately. As expected (A16), the gaseous halo looses mass while the stellar
disc gains it. In Fig. \ref{J_all} we further plot the angular momentum as a function of time for these three components, and we can see that the halo gas is gradually loosing angular
momentum over time, while the angular momentum of the stellar disc is increasing.  The gas in the disc is loosing mass and angular momentum due to star formation in the disc. These plot thus show how the gas in the halo is accreted onto the
disc, where it forms stars (see A16). Therefore, the stellar disc is gaining angular momentum by gas
accretion and star formation.  

Note that this simple angular momentum redistribution picture implicitly assumes
that our system is relatively isolated. Otherwise, it will be able to
exchange matter and/or angular momentum with other galaxies, so that
the total mass and angular momentum of our system need not be
conserved quantities. \\

\begin{figure}
\centering
\includegraphics[scale=0.3]{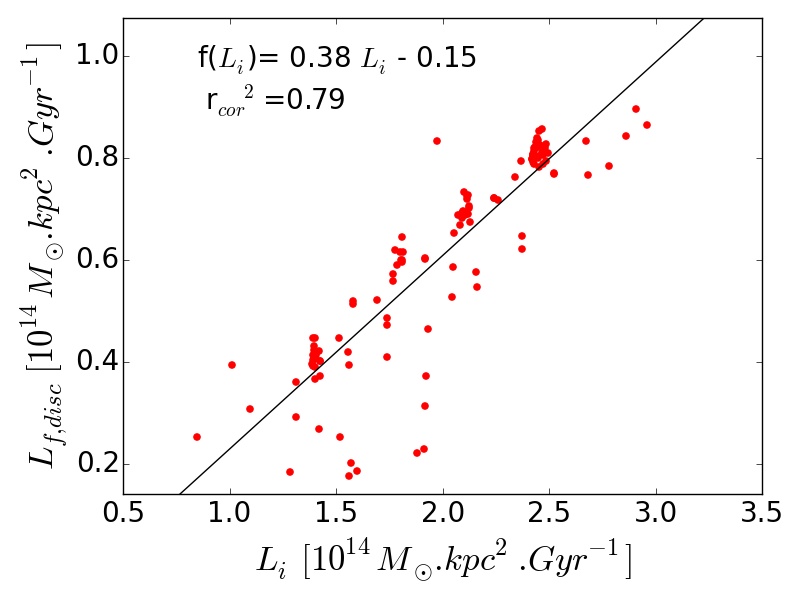}
\caption{Final baryonic disc angular momentum, as a function of the total baryonic angular momentum of the system computed at the merging time, for simulations in sample A.}
\label{JiJf_corr}
\end{figure}

\noindent Thus, the disc grows from material taken from the gaseous halo, which
directly gives a fraction of its angular momentum to the disc. To find how this fraction varies for different simulations, we plot (Fig. \ref{JiJf_corr}) the final baryonic disc angular momentum (as defined in section \ref{sec:bary_disc}) versus the initial total baryonic angular momentum of the system (which in practice we calculate at the merging time). We find a clear correlation, which shows that gaseous haloes with more angular momentum create on average discs with more angular momentum.

To conclude, the initial
properties of the halo -- and in particular its angular momentum --
affect the final disc structure via gas accretion, which constitutes a plausible
reason for the disc properties (scalelengths and break radius) to be
linked to the initial spin parameter $\lambda$. 

\section{Discussion}
\label{sec:discussion}

\subsection{Robustness of the correlations}
\label{sec:discu_gen}

\begin{figure*}
\centering
\includegraphics[scale=0.2]{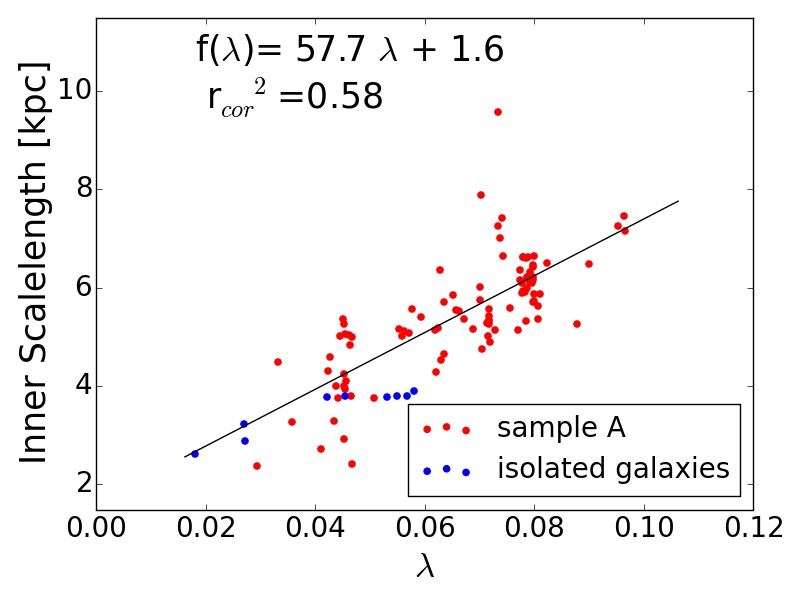}
\includegraphics[scale=0.2]{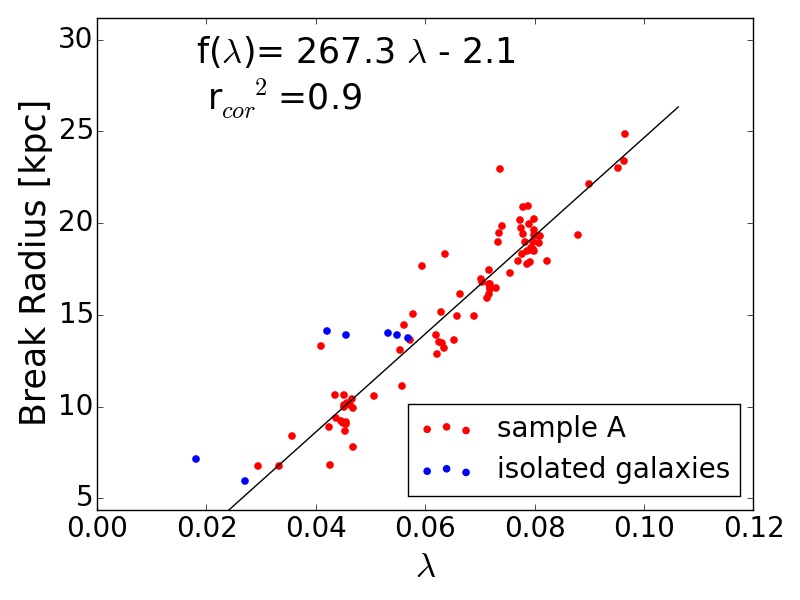}
\includegraphics[scale=0.2]{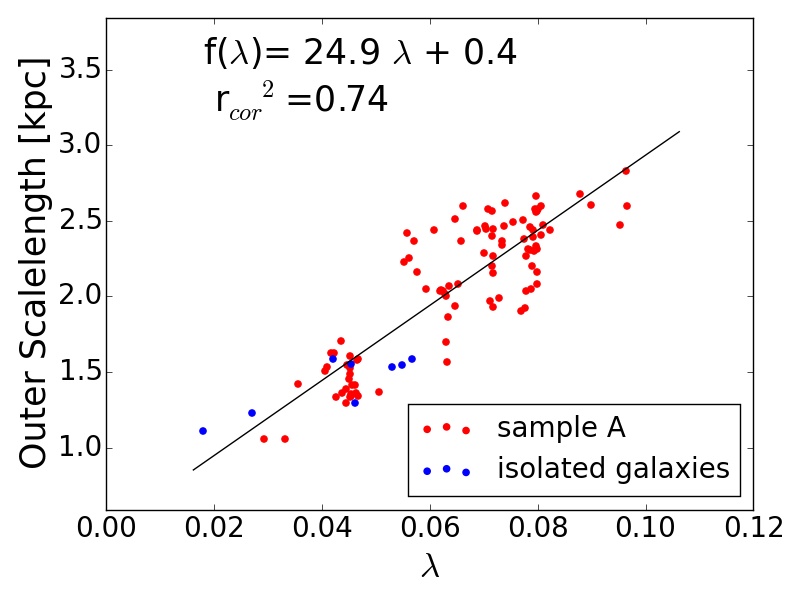}
\caption{Same plots as in Fig. \ref{res}, but adding simulations of isolated galaxies (in blue).}
\label{idfmdf}
\end{figure*}

\begin{figure*}
\centering
\includegraphics[scale=0.2]{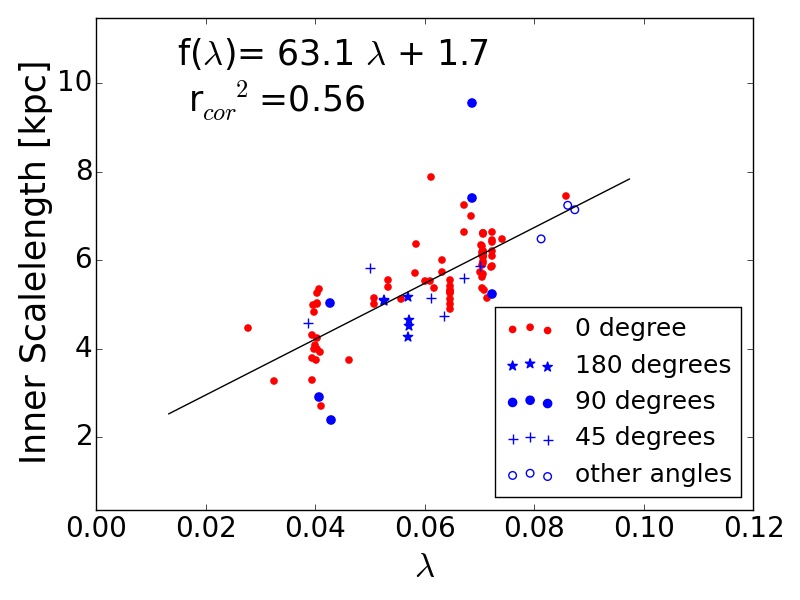}
\includegraphics[scale=0.2]{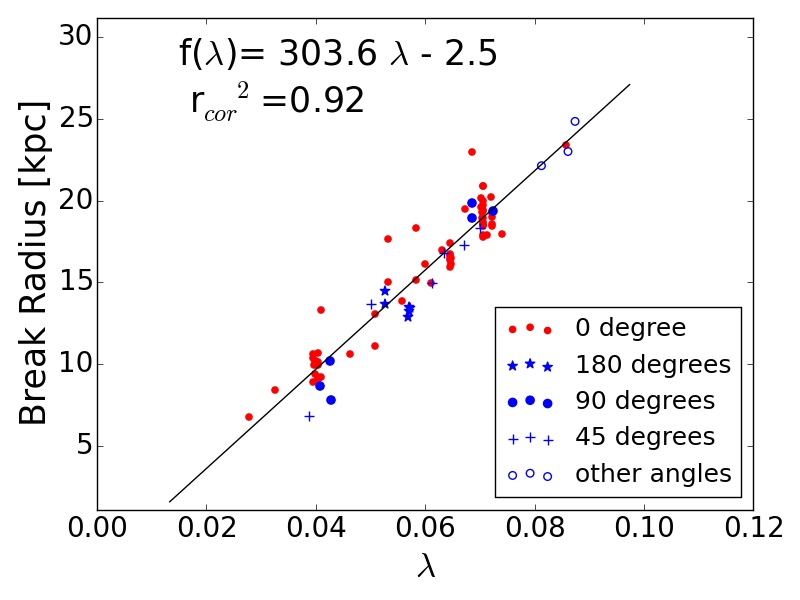}
\includegraphics[scale=0.2]{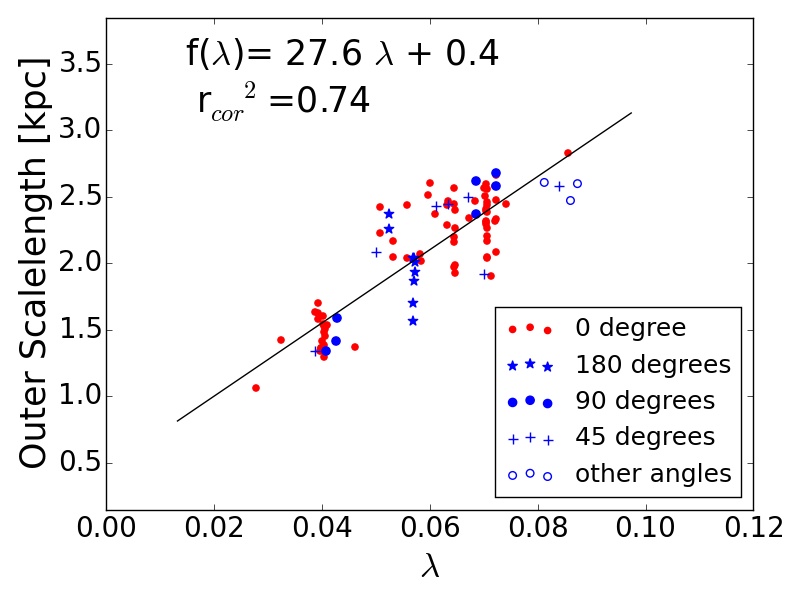}
\caption{Same plots as in Fig. \ref{res}, but showing the effect of
  spin orientation on our correlations, with the angle between the two protogalaxies axes.}
\label{res_angles}
\end{figure*}

To test the robustness of our results with respect to the time
defined as final state, we fitted the radial density profiles also 5.8 Gyr after the merging,
instead of 7.8 Gyr. The values of the correlation
coefficients are reported in Table \ref{table1} (2nd row).  As expected due to the growth of the disc, the individual values of the scalelengths and the break radii
are different, but we still find increasing linear trends with
$\lambda$. Therefore, the time chosen
as final state does not seem to be important for this study, provided it is consistent for all simulations and provided it is sufficiently long after the merging for the disc to have settled. Under these conditions, the increasing trend
with $\lambda$ should be valid regardless of the disc evolution time. We can
thus keep 7.8 Gyr after the merging as the final state without being concerned about the effect of our choice.

We also changed the time at which we calculate $\lambda$
(the initial state) and used the merging time. We found very little
difference in the results, and the correlation coefficients are
similar (Table \ref{table1}, 3rd row). We repeated this analysis, calculating $\lambda$
at various times covering the range
between the start of the simulation and the merging, and conclude
that the time chosen does not change the results significantly. We can
also use $\lambda$ at the final state instead of the initial state, to link $\lambda$ and the disc structure both taken at the same time. We found again the same increasing trends, with the correlation
coefficients reported in Table \ref{table1} (4th row). This points out the existence of a correlation between
the initial and final $\lambda$, which we confirmed by plotting one
against the other, and argues that for an
analysis of this type, one can take the value of
angular momentum at any time, as long as it is consistent for all
simulations.

We also used $\lambda$ at t=0 Gyr for all
simulations (sample A), and again find correlations for the scalelengths and the break
radius, although the correlation coefficients are slightly lower (Table \ref{table1}, last row). \\

\noindent To see if the results obtained for our merger simulations are
also valid for isolated galaxies, we included 12
simulations which take only one of the protogalaxies used in the merger simulations, and see how
it evolves in isolation. This protogalaxy contains the same number of particles as the simulations of sample A, to keep the same total mass. To be able to compare the merger to the isolated
simulations, we define the initial state as the start of the new disc
formation for both, i.e. at the end of the merging for the former, and at t=0 for the latter. We take 7 Gyr after
the initial state as the final state. We find that the
isolated galaxies fit well with the merger simulations (see
Fig. \ref{idfmdf}), which argues that the increasing linear trends of
the two scalelengths and the break radius with the angular momentum
seem to be the same if the disc is formed in isolation or from a major
merger. Our results should therefore also be valid for isolated galaxies. \\

\noindent \cite{Herpich.SD.15} did a similar analysis to ours, looking at the
dependence of the scalelengths and the break radii with the angular
momentum, but using a sample of 9 simulations of galaxies evolved in
isolation. They computed the spin parameter $\lambda$ of the halo at
the start of the simulation, and also found that the inner scalelength
and the break radius increase with $\lambda$. However, they observed a decreasing trend of the outer
scalelength with $\lambda$, while we have an increasing one. We
explain this difference by the fact that they included upbending
discs in their analysis, whose outer scalelengths are naturally much higher
than for downbending discs, while we only have downbending discs in
our sample. Upbending profiles will be discussed in detail in a future paper
(Paper IV). As a consequence, \cite{Herpich.SD.15} find an
increasing trend with $\lambda$ for the ratio of the inner scalelength 
to the outer scalelength, while we do not, our
values showing no correlation with $\lambda$. \\

\begin{table*}
  \caption{Parameters of the linear fits for different plots of $\lambda$
    versus the inner
    scalelength, break radius and outer scalelength. $t_i$, $t_f$
    and $t_{merg}$ are respectively the times of the initial state,
    the final state and the merging, and $r_c^2$ is the correlation
    coefficient for the linear fit with equation: $f(\lambda)=a\lambda+b$.}
\label{table1}
  \centering
  \begin{tabular}{|c|c|c|c||c|c|c|c||c|c|}
    
    \hline
    & \multicolumn{3}{|c|}{Inner Scalelength} &
    \multicolumn{3}{|c|}{Break Radius} &
    \multicolumn{3}{|c|}{Outer Scalelength}\\
    & $r_c^2$ & $a$ & $b$  & $r_c^2$ & $a$ & $b$  & $r_c^2$ & $a$ & $b$ \\ 
    \hline
    $\lambda_{t=t_i}$, $t_f=t_{merg}+7.8$ & 0.56 & 63.1 & 1.7 & 0.92 & 303.6 & -2.5 & 0.74
    & 27.6 & 0.4 \\
    \hline
    $\lambda_{t=t_i}$, $t_f=t_{merg}+5.8$ & 0.48 & 55.0 & 2.3 & 0.91 & 283.5 & -1.9 &
    0.74 & 24.8 & 0.4\\
    \hline
    $\lambda_{t=t_{merg}}$, $t_f=t_{merg}+7.8$ & 0.56 & 58.0 & 1.7 & 0.90 & 277.9 & -2.2 & 0.72 &
    24.9 & 0.5\\
    \hline
    $\lambda_{t=t_{f}}$, $t_f=t_{merg}+7.8$ & 0.55 & 54.0 & 1.6 & 0.90 & 259.8 & -3.0 & 0.73 & 23.7 & 0.4\\
    \hline
    $\lambda_{t=t_f}$, $t_f=t_{merg}+5.8$ & 0.47 & 47.0 & 2.2 & 0.87 & 238.6 & -1.8 &
    0.71 & 20.9 & 0.4\\
    \hline
    $\lambda_{t=0}$, $t_f=t_{merg}+7.8$  & 0.53 & 59.8 & 2.0 & 0.89 & 291.1 & -1.1 &
    0.72 & 26.6 & 0.6\\
    \hline
  \end{tabular}
\end{table*}

\begin{figure*}
\centering
\includegraphics[scale=0.2]{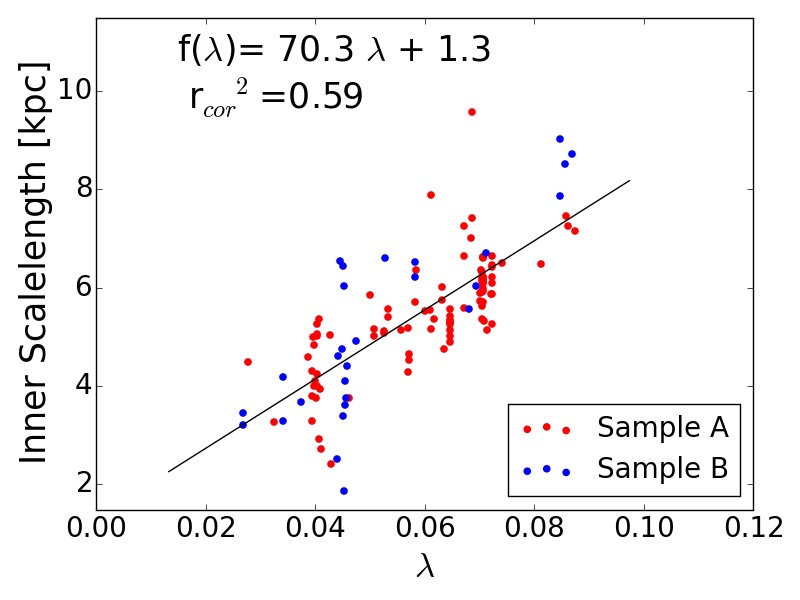}
\includegraphics[scale=0.2]{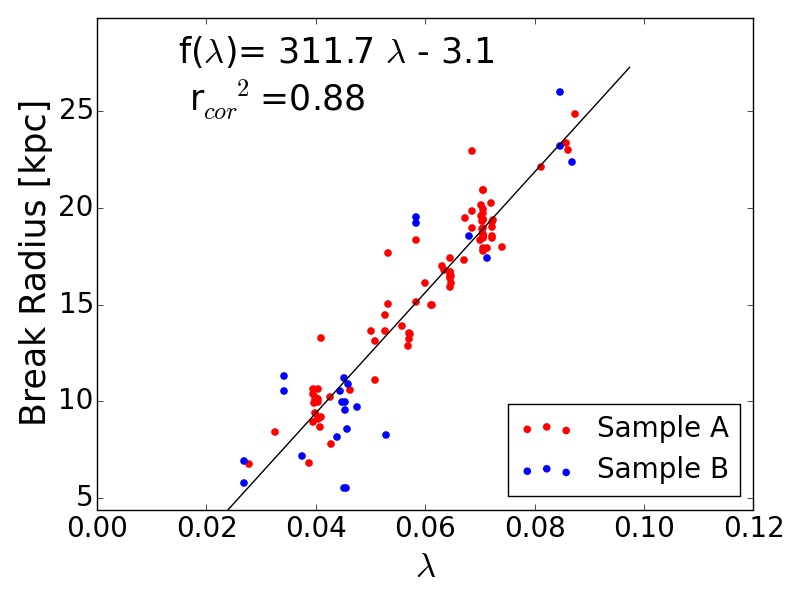}
\includegraphics[scale=0.2]{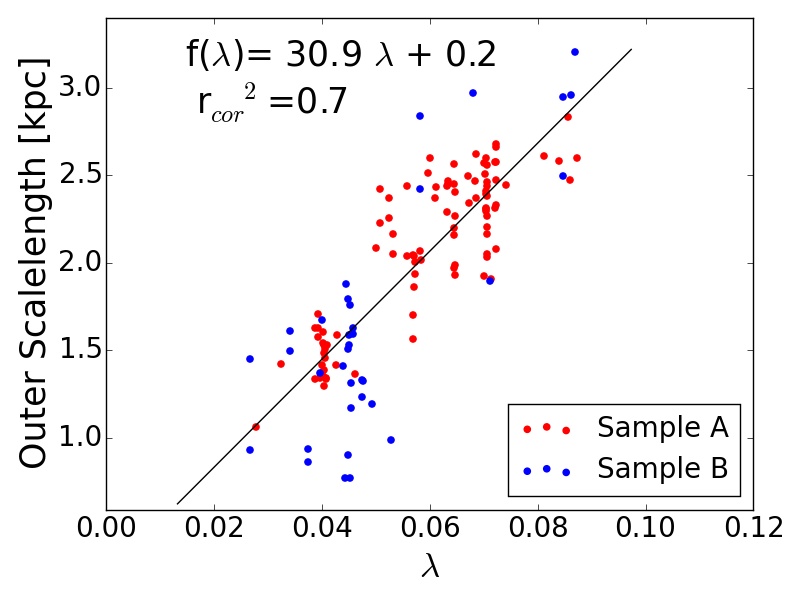}
\caption{Same plots as in Fig. \ref{res}, but adding a sample of 67 simulations (sample B) with total masses different from those of the simulations in sample A.}
\label{res_mass}
\end{figure*}

\noindent We mentioned in section \ref{sec:desc_simus} that in some
simulations, the spin axis of one or both protogalaxies is tilted by a given angle; this concerns 34 simulations in sample A. This angle can be around the X
or the Y axis, Z being the axis perpendicular to the orbital plane. We
wanted to see if
these simulations behave differently in our results than the simulations where both protogalactic
spin axes are parallel, and perpendicular to the orbital plane. In
Fig. \ref{res_angles} we plotted again our correlations between the
scalelengths and $\lambda$, but showing the effect of spin orientation. We can see that the simulations
where the spin axes are not parallel (angle $\neq$ 0) fit reasonably well with the
others. We
can thus conclude that our results seem to be valid regardless of the
spin axis orientation.

We also looked at the effect of the presence of a central AGN on our
correlations, and found that the simulations without AGN also fit well
with the ones having an AGN. This confirms that our AGN only affects
the central part of the galaxy (R17).

The effect of the initial density distribution of gas and DM have also been tested since our sample contains simulations with different distribution parameters, such as the initial characteristic radii of the halo (see A16) or the presence of a central core for DM. Although the corresponding parameter space is very large, our relatively few trials argue that such  parameters do not seem to have an effect on the correlations we found. Various other parameters, such as the merging orbit, the baryonic to total mass ratio, and the softening of the gas and DM, were also shown to have no significant impact on our results.

\subsection{Introducing simulations with different total masses}
\label{sec:diffmtot}

Here we will analyse the effect of changing the total mass in the simulations, using sample B (see section \ref{sec:desc_simus}).
The problem when analysing simulations of different masses is that the evolution time-scale is also different. In our analysis we took a fixed evolution time after the merging (7.8 Gyr) to compare the properties of the disc, which introduces a small inconsistency when we compare simulations with different total masses.

Nevertheless, we expect the time-scales to be similar if the total masses are not too different, which is the case of the simulations in our subsample of 67 galaxies (sample B) since they have total masses between 0.5 and 2 times the total mass of simulations in sample A. We include them in our correlations (Fig. \ref{res_mass}) and find again high correlation coefficients; the sample B fits well with sample A. Furthermore, we added simulations of galaxies evolved in isolation as in section \ref{sec:discu_gen}, but with half the number of particles, i.e. with a mass twice lower than the simulations of sample A. We found that these simulations match well in the correlations, which again suggests that our results are valid also for galaxies formed in isolation. 
Moreover, we included the simulations of sample B in the correlation of the final baryonic disc angular momentum as a function of the initial baryonic angular momentum (as in Fig. \ref{JiJf_corr}), and found again a very good match.

Since the simulations of sample B follow similar trends as sample A despite their total masses being different, in the following subsections we will include them in our analysis, using the A+B sample.

\subsection{Migration and angular momentum}
\label{sec:mig}

The formation of outer discs in downbending profile galaxies is still debated, but one of the main possible scenario relies on the presence of  
outwards stellar migration to build the outer dic (\citealt{Roskar.DSQ.08}). We thus investigate where 
the stars in the outer disc of our remnant galaxies come from, by calculating the radius at which
they were born. The stellar discs of the two protogalaxies are
separate before the collision and thus the birth radius 
with respect to the remnant galaxy cannot be defined. We 
therefore use in our analysis only the stars born after the merging, as in section \ref{sec:fit}.

\begin{figure}
\centering
\includegraphics[scale=0.4]{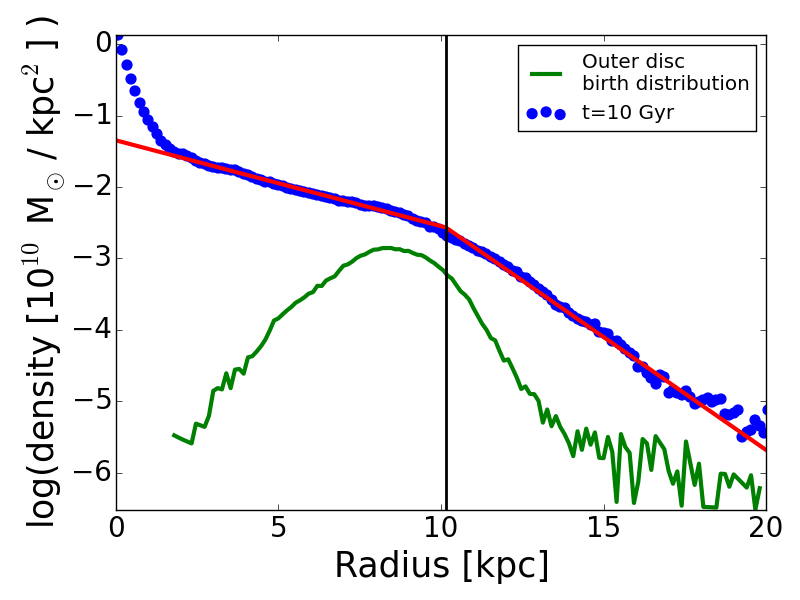}
\caption{Radial density profile for a snapshot at t=10 Gyr (in blue),
and distribution at time of birth of the stars ending up in the outer
disc (in green). The fit of the profile at t=10 Gyr is plotted in red, and the corresponding break radius is indicated with a black vertical line.}
\label{migout}
\end{figure}

We define at any given time the stars of the outer disc as all the stellar particles located
beyond the break. We plot in Fig. \ref{migout}
the distribution of the radii at birth of the stars which by t=10 Gyr
end up in the outer disc, for simulation $mdf730$ from sample A, which was chosen because of its smooth profile, allowing us to make our points on migration clearer. Since every star is born 
at a different time, this distribution does not represent a
real stellar distribution at any time of the simulation, but it helps visualise
the fraction of stars born at a given radius throughout the
whole simulation. We can see that most stars of the outer disc ($78\pm18$ per cent for sample A+B) were born at
a radius less than the break radius found at t=10 Gyr. This means that these stars have migrated radially outwards to form the outer disc, establishing the presence of inside-out
stellar migration in the disc. This is in agreement with the results
of \cite{Roskar.DSQ.08} presenting simulations with a downbending
disc profile induced by a star formation threshold and outward
migration, and with observations e.g. in \cite{Radburn-Smith.RDD.12}
and \cite{Zheng.THM.15}, and thus adds a further argument to those
presented in A16 that the remnant of a major merger is a disc galaxy
and behaves in all aspects as such. 

\begin{figure}
\includegraphics[scale=0.3]{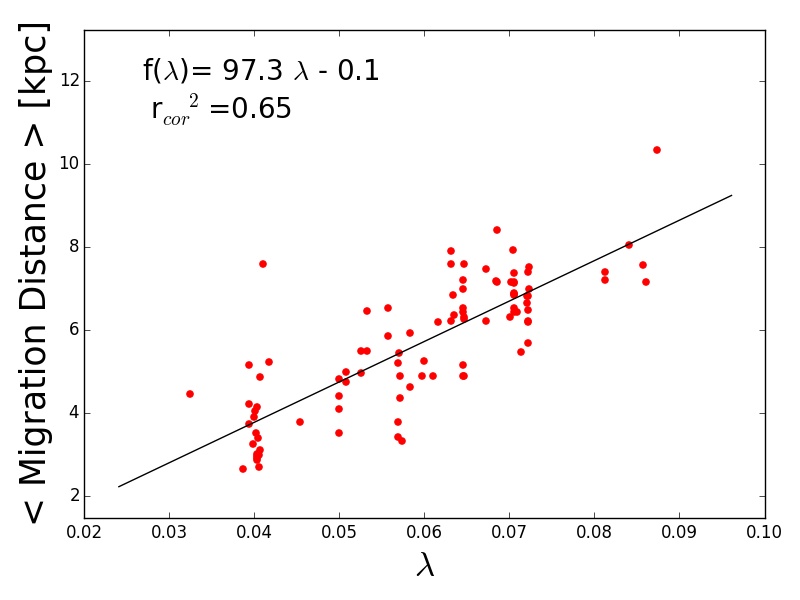}
\caption{Mean inside-out migration distance of the stars ending up in
  the outer disc, as a
  function of the initial spin parameter $\lambda$, for sample A+B.}
\label{distmig_lambda}
\end{figure}

Having established that there is inside-out stellar migration
  in our discs, we explore the effect of angular momentum on this
  phenomenon. To characterise the migration, we calculate the distances
  travelled by these stars migrating towards the outer disc,
  throughout the whole simulation. To derive the migration distances
for a given simulation, we first compute
the radial distance travelled by each star ending up in the outer disc
between its time of birth and the final state (as defined in section \ref{sec:fit}):

\begin{equation}
\centering
D_{mig} = R_{final}-R_{birth},
\end{equation}

\noindent where R$_{final}$ and R$_{birth}$ are the radii at the final
state and at birth for each star. We remove again the
stars born before the merging since migration only starts after the collision, or, more precisely, after the disc formation begins (t$_{bd}$, A16). Since we are only interested in the outward
migration towards the outer disc, we also remove the stars migrating
inwards, which represent only about 5 per cent of the outer disc population.
We then take the mean over all the remaining stars to get a single value for
each simulation, which we plot versus $\lambda$.
We repeat this for all the simulations in sample A+B and plot the result in Fig.
\ref{distmig_lambda}. We see a clear increasing trend of the migration
distances with $\lambda$, allowing us to conclude that higher initial
angular momentum systems lead to discs with higher migration
distances towards the outer disc. In fact, we find the same result if
we consider inside-out migration in the whole disc, and do not restrict
ourselves to outer disc stars. The stars travel globally further out in high
angular momentum systems, and this is particularly important in the
outer disc, where 94 ($\pm$5) per cent of the stars have undergone inside-out
migration (compared to 46$\pm$5 per cent for the whole disc).

The angular momentum thus seems to play an important role in the
radial outward migration, by redistributing the stellar content in the outer
disc efficiently (with large travelling distances) or
not.

This result is consistent with outer discs in high $\lambda$ simulations having larger scalelengths: if stars travel further out, their final distribution will be more extended, resulting in a larger outer disc.

\subsection{Migration and birth radius}
\label{sec:migbirth}

In section \ref{sec:mig}, we showed that most particles ending up in the outer
disc were born inside the break radius, as calculated at t=10 Gyr. However, the stars whose birth distribution has been
plotted in Fig. \ref{migout} were born at all times after the merging. As will be shown for our simulations in Paper IV (see also \citealt{Perez.04} and Azzollini et al. 2008 for observations), the break radius generally increases with time, so that
stars which are born inside the 10 Gyr break in Fig. \ref{migout}
could in fact be born beyond the break taken at their time of
birth. Therefore, we cannot claim that about 80 per cent of the outer disc's stars
that were born at a radius which corresponds to the inner disc at
10 Gyr were actually born in the inner disc, since their birth radius
could be in the outer disc at their birth time. Although
this does not change the fact that most of these stars clearly migrated
outwards, it could question the claim that they travelled from the inner
disc to the outer disc.

 To solve this, we took the same stars (i.e. the
ones in the outer disc at 10 Gyr), and computed the fraction which was born in the inner disc defined with respect to the
break at their time of birth. This fraction however depends
strongly on the value taken for the break radius, since the break is often
located near the end of structures such as spirals (\citealt{Laine.LS.14}; Paper IV), which are high star formation areas (A16 for
simulations and e.g. \citealt{Silva-Villa.Larsen.12} for observations). Besides, the
break radius can be difficult to measure precisely at early times (a few Gyr after the
merging). We thus defined errorbars for the estimated break position
at each time step for every simulation, and
excluded all the particles born closer to the break radius than these errorbars. To make sure
our 10 Gyr outer disc definition is also reliable, we
defined the particles of the outer disc as the ones located beyond
the external errorbar of the break radius.

To exclude the effect of radial displacements due to radial oscillations of stars
in their orbit, we chose to use $R_G$ -- the guiding radius of each
star -- instead of their regular radius to derive their position with
respect to the break. $R_G$ is the radius of the
circular orbit associated to the angular momentum of a given star, and
can be derived by solving the equation:

\begin{equation}
\centering
R_{G}=\frac{L_z}{v_{circ}(R_{G})},
\end{equation}

\noindent where $L_z$ is the vertical component of the angular momentum of the star, and $v_{circ}$ is the
circular velocity at the guiding radius.

The fractions of outer disc stars that are born in the inner disc derived with this method
cover a wide range of values depending on the simulations, from 10 to 80 per cent. Nevertheless, we find globally lower
fractions than the ones from section \ref{sec:mig}, as well as the ones from \citet[$\sim$ 85 per cent]{Roskar.DSQ.08}, which can be
explained by the fact that the break radius tends to increase with
time (see paper IV), so that some stars were born inside the 10 Gyr break
but outside their birth time break, and thus should be considered as being born in the outer
disc. Note that these low fractions can not be explained by our choice to take the guiding radius instead of the regular radius, or to exclude the stars within the errorbars of the break radius: we performed the same analysis using the radius instead of the guiding radius, and also keeping the particles inside the errorbars, and found similar results in both cases.

\begin{figure}
\centering
\includegraphics[scale=0.28]{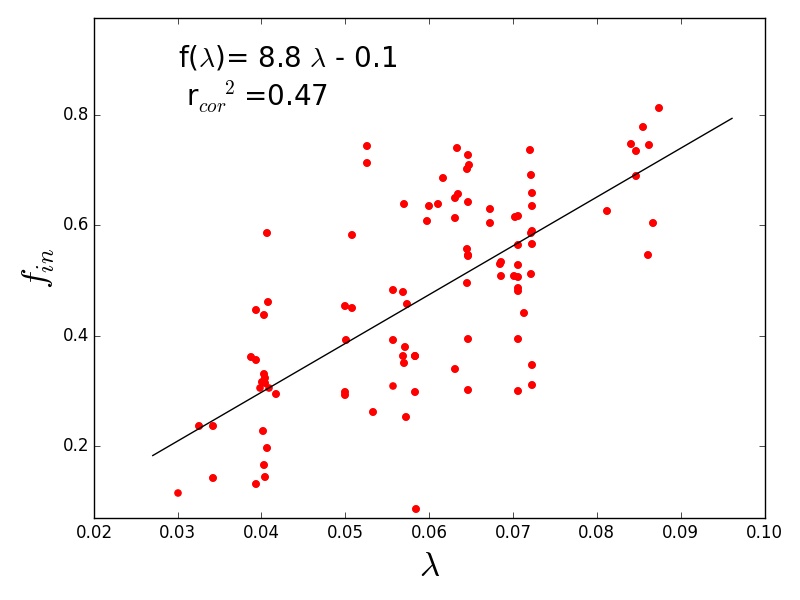}
\caption{Fraction of outer disc stars which were born in the inner
  disc, as a function of the spin parameter $\lambda$, for sample A+B. For each star, the inner disc is defined
  here with respect to the
  break radius derived at its time of birth, as explained in section \ref{sec:migbirth}.}
\label{frac}
\end{figure}

To understand the spread of values derived for the fraction of
stars born in the inner disc, we tried to plot these fractions again versus
the spin parameter $\lambda$ (taken for all the components, 1 Gyr before the
merging) in Fig. \ref{frac}. Although having a large scatter
(probably because of the sensitivity of the values to the break
location and its errorbars, and the difficulty to define them precisely at early times), this
plot shows again an increasing trend of the fractions with
$\lambda$. Therefore, galaxies with high angular momentum will have an outer disc
which is populated by a majority of stars coming from the inner disc,
whereas in low angular momentum galaxies, stars in the outer disc will
mostly be born in the outer disc.
It is important to note that the trend found in Fig. \ref{frac} involves the fraction of stars, i.e. relative and not absolute numbers. It thus does not imply that low angular momentum galaxies form more stars in
their outer parts, since the number of stars in the outer disc is
different in every simulation. \\

We thus have simulations which have outer discs formed mostly by
radial migration, but with a significant amount of stars in the outer parts which were born in
the outer disc. We conclude that not taking into account the fact 
that the break radius changes with time, can lead to overestimating the fraction 
of stars born in the inner disc in the context of radial
migration. Therefore, the picture of outer disc stars mostly coming from the
inner disc has to be used with caution, since it depends on the angular momentum of the galaxy. Only galaxies with high spin parameter $\lambda$ ($\lambda$>0.08 in our simulations) seem to be
able to have about 80 per cent of their outer disc stars born in the
inner disc. In galaxies with average or yet lower angular momentum
($\lambda$<0.06 in our simulations) this fraction drops below 50 per cent,
disclaiming the inner disc as the main origin of the outer disc stars.

\section{Summary and conclusion}
\label{sec:ccl}

For this series of papers we ran a sample of high resolution simulations of which three fiducial cases were described in A16, where more general information on the runs was also given. It was shown there that after a merging, a 
new disc forms in the remnant from gas accreting from the halo. In this paper, we used subsamples derived as described in section \ref{sec:desc_simus}. We derived the stellar radial density profiles of the remnants at the end of all the
simulations, and found downbending (type II) profiles. We then derived the corresponding inner and outer disc scalelengths, as well
as the break radius. We used the spin parameter $\lambda$ computed 1 Gyr before the merging to have  
a definition of the angular momentum consistent for all our
simulations. Plotting for our sample the values derived from the radial density profiles fits as a function of $\lambda$, we found that the inner, the outer scalelength and the break radius increase with $\lambda$. Therefore, both the inner and the outer discs are larger for
higher angular momentum systems. \\ 
 
\noindent To explain how the initial orbit- and halo-dominated spin parameter $\lambda$ of the merging system can affect the properties of the final remnant disc (mostly composed  
of stars born after the merging), we looked at the angular momentum
redistribution in our simulations. The scalelengths and break radius correlate with the baryonic angular momentum of the disc, suggesting a link between the latter and the initial angular momentum. To understand how the angular momentum of the disc was acquired, we investigated the transfers of the baryonic matter and angular momentum between the halo and the disc, after making sure that the total baryonic angular momentum is conserved. In the framework of disc galaxy formation from a major merger of two progenitors with extended gas-rich haloes (A16), it is easy to understand that the gaseous halo gives angular momentum to the disc by accreting its gas onto it. We showed this in two fiducial simulations with different $\lambda$ values, with the halo gas gradually loosing a fraction of its mass and angular momentum to the stellar disc, by gas accretion and star formation.
We further found that haloes with higher initial total angular momentum (or $\lambda$) will create final discs with higher angular momentum, and larger scalelengths and break radius.

Naturally this scenario is only valid for
an isolated system with no
interactions with the environment. Thus our results, as well as those of other dynamical (and thus necessarily idealised) simulations, can not be straightforwadly extended to a cosmological context, since interactions with environment could interfere with the angular momentum exchanges between the gaseous halo and the baryonic disc.\\

\noindent The correlations of the scalelengths and the break radius with $\lambda$ are robust and do not change with 
the time at which we fit the radial density profiles, or the time used to compute $\lambda$. Furthermore, simulations of isolated galaxies follow the same  
trends, so that our results should be valid 
for discs formed both in isolation and in major mergers. In some 
simulations, the spin axes of the protogalaxies are tilted, and we showed that the spin axis orientations of the two merging 
protogalaxies do not seem to play a role in this analysis.

We also included in the correlations a sample of 67 merger simulations with total masses lower or higher than the simulations from the main sample. Although the evolution time-scale of these simulations is also different, we found that these simulations (which have total masses reasonably close to our main sample) fit well with the others in the correlations, and thus added them to the sample.
 
We analysed the outer disc origin, and found that the stars ending up in the outer disc were 
mostly born at smaller radii ($\sim$95 per cent), suggesting inside-out migration as the 
main formation driver for the outer disc, in good agreement with previous work on galaxies formed in isolation (\citealt{Roskar.DSQ.08}). To see the effect of the angular momentum on this migration, we computed the distance radially travelled by the stars 
towards the outer disc, and plotted it against the spin parameter $\lambda$ for the simulations of our sample. We found a clear correlation, galaxies with higher angular 
momentum having larger inside-out migration distances.

We also showed that to study the origin of the outer disc, it is necessary to take into account the fact that 
the break location changes with time, 
so as to avoid overestimations of the fraction of stars born in the inner 
disc. While $\sim$80 per cent of the outer disc stars were born inside the 
break derived at t=10 Gyr, this fraction can take lower values (under 50 per cent) using the 
time dependent break, which sets doubt on the picture of the outer disc formed mainly from the inner 
disc. This fraction depends on the angular momentum, and is higher in high angular momentum galaxiess. Thus, in some low angular momentum systems, this fraction can drop to values even lower than 20 per cent.\\ 
 
\noindent We can thus conclude that the angular momentum is a key parameter in the 
creation of disc structures, as it affects radial migration, and can explain the large range of 
values observed for the inner and outer scalelengths in disc 
galaxies of a given mass.

\vspace{1cm}
\textbf{ACKNOWLEDGEMENTS}
\vspace{0.3cm}

\noindent We thank Jean-Charles Lambert for computer assistance, and the referee for useful suggestions. This work was supported in part by the Polish National Science Centre under grant 2013/10/A/ST9/00023, and was granted access to the French HPC resources of
[TGCC/CINES/IDRIS] under the allocations
2014-[x2014047098], 2015-[x2015047098] and 2016-[x2016047665], made
by GENCI, as well as the HPC resources of Aix-Marseille Universit\'e financed by the project Equip@Meso (ANR-10-EQPX-29-01) of the program « Investissements d’Avenir » supervised by the Agence Nationale de la Recherche.

\vspace{1cm}

\end{document}